\documentclass[english,10pt,twocolumn,twoside]{IEEEtran}
\usepackage[T1]{fontenc}
\usepackage[latin9]{inputenc}
\usepackage{bm}
\usepackage{amsmath}
\usepackage{amssymb}
\usepackage{graphicx}

\makeatletter

\providecommand{\tabularnewline}{\\}

\usepackage{colortbl}
 \usepackage{cite}

\makeatother

\usepackage{babel}
\begin{document}

\title{Generalized Rao Test for Decentralized Detection of an Uncooperative
Target}

\author{D.~Ciuonzo,~\IEEEmembership{Senior~Member,~IEEE}, P. Salvo~Rossi,~\IEEEmembership{Senior~Member,~IEEE}
and P. Willett,~\IEEEmembership{Fellow,~IEEE}\thanks{Manuscript received 26th November 2016; revised 1st March 2017; accepted
10th March 2017. The associate editor coordinating the review of this
manuscript and approving it for publication was Dr. F. Kamalabadi.\protect \\
D. Ciuonzo is with Network Measurement and Monitoring (NM-2) s.r.l.,
Naples, Italy (e-mail: domenico.ciuonzo@ieee.org).\protect \\
P. Salvo Rossi is with the Dept. of of Electronic Systems, NTNU, Trondheim,
Norway. (e-mail: salvorossi@ieee.org).\protect \\
P. Willett is with the Dept. of Electrical and Computer Engineering,
University of Connecticut, Storrs (CT), US (e-mail: willett@uconn.edu).}\vspace{-0.2cm}
}
\maketitle
\begin{abstract}
We tackle distributed detection of a non-cooperative target with a
Wireless Sensor Network (WSN). When the target is present, sensors
observe an (unknown) deterministic signal with attenuation depending
on the distance between the sensor and the (unknown) target positions,
embedded in symmetric and unimodal noise. The Fusion Center (FC) receives
quantized sensor observations through error-prone Binary Symmetric
Channels (BSCs) and is in charge of performing a more-accurate global
decision. The resulting problem is a two-sided parameter testing with
nuisance parameters (i.e. the target position) present only under
the alternative hypothesis. After introducing the Generalized Likelihood
Ratio Test (GLRT) for the problem, we develop a novel fusion rule
corresponding to a Generalized Rao (G-Rao) test, based on Davies'
framework, to reduce the computational complexity. Also, a rationale
for threshold-optimization is proposed and confirmed by simulations.
Finally, the aforementioned rules are compared in terms of performance
and computational complexity.
\end{abstract}


\section{Introduction\label{sec: Introduction}}

\IEEEPARstart{W}{ireless} Sensor Networks (WSNs) have attracted significant
interest due to their applicability to reconnaissance, surveillance,
security and environmental monitoring \cite{Chong2003}. Distributed
detection is one of the main tasks for a WSN and it has been heavily
investigated in the last decades \cite{Varshney1996}.

Due to stringent bandwidth and energy constraints, it is often assumed
that each sensor sends one bit of information about the estimated
hypothesis to the Fusion Center (FC). In this context the optimal
test (under Bayesian/Neyman-Pearson frameworks) at each sensor is
known to be a one-bit quantization of the local Likelihood-Ratio (LR);
that is to perform a LR Test (LRT). Unfortunately in most cases, due
to a lack of knowledge of the parameters of the target to be detected,
it is not possible to compute the LRT at each sensor. Also, even when
the sensors \emph{can} compute their local LRT, the search for local
quantization thresholds is exponentially complex \cite{Tsitsiklis1993,Viswanathan1997}.
Thus the bit of information being sent is usually the result of a
``dumb'' quantization \cite{Ciuonzo2013b,Fang2013} or represents
the estimated binary event, according to a sub-optimal rule \cite{Ciuonzo2014a,Ciuonzo2015a}.
In both cases, the bits from the sensors are collected by the FC and
combined via a specifically-designed fusion rule aiming at improved
detection rate.

The optimum strategy to fuse the sensors' bits at the FC, under conditional
independence assumption, is a weighted sum, with weights depending
on unknown target parameters \cite{Varshney1996}. Some simple fusion
approaches, based on the counting rule or channel-aware statistics,
have been proposed in the literature to overcome such unavailability
\cite{Aalo1994,Chen2004,Niu2008,Ciuonzo2012}. On the other hand,
in some particular scenarios the uniformly most powerful test is independent
of the unknown parameters under the alternative hypothesis, so they
do not need to be estimated \cite{Ciuonzo2013}. Nonetheless, in the
general case the FC is usually in charge of solving a composite hypothesis
test and the Generalized LRT (GLRT) is commonly employed \cite{Kay1998}.
Indeed, GLRT-based fusion of quantized data was studied in \cite{Niu2006b,Fang2013,Iyengar2012}
for: ($i$) detecting a known source with unknown location, ($ii$)
detecting an unknown source with known observation coefficients, and
$(iii)$ fusing conditionally dependent decisions, respectively. As
a simpler alternative, a Rao test was developed in a more general
context for problem $(ii)$ in \cite{Ciuonzo2013b}. However, in the
case of an \emph{uncooperative} \emph{target}, it is reasonable to
assume that both the target emitted signal and location are not available
at the FC. To the best of authors' knowledge, only a few works have
dealt with the latter case \cite{Shoari2012,Ciuonzo2016}. In \cite{Shoari2012},
a GLRT was derived for revealing a target with unknown position and
emitted power and compared to the so-called counting rule, the optimum
rule and a GLRT based on the awareness of target emitted power, showing
a marginal loss of the latter rule with respect to the ``power-clairvoyant''
GLRT. Unfortunately, the considered GLRT requires a grid search on
both the target location and emitted power domains. Therefore, as
a computationally simpler solution, generalized forms of locally-optimum
detectors have been proposed for non-cooperative detection of a fluctuating
target emission \cite{Ciuonzo2016}.

In this letter, we focus on decentralized detection of a non-cooperative
target with a spatially-dependent emission (signature), with emitted
signal modelled as unknown and deterministic (as opposed to \cite{Ciuonzo2016}).
More specifically, the received signal at each individual sensor is
embedded in unimodal zero-mean additive noise, with a deterministic
Amplitude Attenuation Function (AAF) depending on the sensor-target
distance. Each sensor observes a local measurement on the absence/presence
of the target and forwards a single bit version to a FC, over noisy
imperfect (modelled as Binary Symmetric Channels, BSCs) reporting
channels, which is in charge of providing an accurate global decision.
The problem considered is a two-sided parameter test with nuisance
parameters present only under the alternative hypothesis, which thus
precludes the application of conventional score-based tests, such
as the Rao test. In order to reduce the computational complexity required
by the GLRT, we develop a (simpler) sub-optimal fusion rule based
on a generalization of the Rao test \cite{Kay1998}. The aforementioned
detector is also compared in terms of computational complexity. Finally,
simulation results are provided to compare these rules in some practical
scenarios.

The letter is organized as follows: Sec.~\ref{sec: System model}
describes the system model; Sec.~\ref{sec:Rao Test} develops the
generalized form of Rao test and tackles the quantizer optimization
problem, with results validated in Sec.~\ref{sec: Simulation Results}.
Finally, conclusions are in Sec.~\ref{sec: Conclusions}\footnote{\emph{Notation} - Lower-case bold letters denote vectors, with $a_{n}$
being the $n$th element of $\bm{a}$; upper-case calligraphic letters,
e.g. $\mathcal{A}$, denote finite sets; $\mathbb{E}\{\cdot\}$, $\mathrm{var\{\cdot\}}$
and $(\cdot)^{T}$ denote expectation, variance and transpose, respectively;
$u(\cdot)$ denotes the Heaviside (unit) step function; $P(\cdot)$
and $p(\cdot)$ are used to denote probability mass functions (pmf)
and probability density functions (pdf), respectively, while $P(\cdot|\cdot)$
and $p(\cdot|\cdot)$ their corresponding conditional counterparts;
$\mathcal{N}(\mu,\sigma^{2})$ denotes a Gaussian pdf with mean $\mu$
and variance $\sigma^{2}$; $\chi_{k}^{2}$ (resp. $\chi_{k}^{'2}(\xi)$)
denotes a chi-square (resp. a non-central chi-square) pdf with $k$
degrees of freedom (resp. and non-centrality parameter $\xi$); the
symbols $\sim$ and $\overset{a}{\sim}$ mean \textquotedblleft distributed
as\textquotedblright{} and ``asymptotically distributed as''.}.

\section{System Model\label{sec: System model}}

We consider a binary hypothesis test where a collection of sensors
$k\in\mathcal{K}\triangleq\{1,\ldots,K\}$ are deployed in a surveillance
area to monitor the absence ($\mathcal{H}_{0}$) or presence ($\mathcal{H}_{1}$)
of a target of interest having a partially-specified spatial signature.
The problem can be summarized as follows:
\begin{gather}
\begin{cases}
\mathcal{H}_{0}\quad:\quad & y_{k}=w_{k},\\
\mathcal{H}_{1}\quad:\quad & y_{k}=\theta\,g(\bm{x}_{T},\bm{x}_{k})+w_{k},\qquad k\in\mathcal{K};
\end{cases}\label{eq:binary_test}
\end{gather}
In other terms, when the target is present (i.e. $\mathcal{H}_{1}$),
we assume that its radiated (amplitude) signal $\theta$, modelled
as \emph{unknown} deterministic, is isotropic and experiences (distance-dependent)
path-loss and additive noise, before reaching individual sensors.
In Eq. (\ref{eq:binary_test}), $y_{k}\in\mathbb{R}$ denotes the
$k$th sensor measurement and $w_{k}\in\mathbb{R}$ denotes the noise
Random Variable (RV) with \emph{$\mathbb{E}\{w_{k}\}=0$ }and \emph{unimodal
symmetric} pdf\footnote{Noteworthy examples of such pdfs are the Gaussian, Laplace, Cauchy
and generalized Gaussian distributions with zero mean \cite{Kay1998}.}, denoted with $p_{w_{k}}(\cdot)$ (the RVs $w_{k}$ are assumed mutually
independent). Additionally, $\bm{x}_{T}\in\mathbb{R}^{d}$ denotes
the\emph{ unknown} position of the target, while $\bm{x}_{k}\in\mathbb{R}^{d}$
denotes the \emph{known} $k$th sensor position. Both $\bm{x}_{T}$
and $\bm{x}_{k}$ \emph{uniquely} determine the value of $g(\bm{x}_{T},\bm{x}_{k})$,
generically denoting the AAF\footnote{We remark that the results presented in this letter apply to any suitably
defined AAF modelling the spatial signature of the target/event to
be detected.}.

For example, the measurement $y_{k}$ is distributed under $\mathcal{H}_{0}$
(resp. $\mathcal{H}_{1}$) as $y_{k}\,|\,\mathcal{H}_{0}\sim\mathcal{N}(0,\sigma_{w,k}^{2})$
(resp. $y_{k}\,|\,\mathcal{H}_{1}\sim\mathcal{N}(\theta\,g(\bm{x}_{T},\bm{x}_{k}),\,\sigma_{w,k}^{2})$)
when the noise is modelled as $w_{k}\sim\mathcal{N}(0,\sigma_{w}^{2})$.
Then, to meet stringent bandwidth and energy budgets in WSNs, the
$k$th sensor quantizes\footnote{We restrict our attention to deterministic quantizers for simplicity;
an alternative is the use of stochastic quantizers, however their
analysis falls beyond the scope of this letter.} $y_{k}$ into one bit of information, i.e. $b_{k}\triangleq u\,(y_{k}-\tau_{k})$,
$k\in\mathcal{K}$, where $\tau_{k}$ denotes the quantizer threshold.
The bit $b_{k}$ is sent over a BSC and the FC observes an error-prone
version due to non-ideal transmission, i.e. $\hat{b}_{k}=b_{k}$ (resp.
$\hat{b}_{k}=(1-b_{k})$) with probability $(1-P_{e,k})$ (resp. $P_{e,k}$),
which we collect as $\hat{\bm{b}}\triangleq\left[\begin{array}{ccc}
\hat{b}_{1} & \cdots & \hat{b}_{K}\end{array}\right]^{T}$. Here $P_{e,k}$ denotes the (known) BEP of $k$th link.

We underline that the \emph{unknown} target position $\bm{x}_{T}$
is \emph{observable }(i.e. can be estimated) at the FC \emph{only}
when the signal is present, i.e. $\theta\neq\theta_{0}$ ($\theta_{0}=0$).
Therefore, the problem in Eq. (\ref{eq:binary_test}) refers to a
\emph{two-sided parameter test }(that is $\{\mathcal{H}_{0},\mathcal{H}_{1}\}$
corresponds to $\{\theta=\theta_{0},\theta\neq\theta_{0}\}$)\emph{
with nuisance parameters ($\bm{x}_{T}$) present only under the alternative
hypothesis $\mathcal{H}_{1}$ }\cite{Davies1987}. The aim of this
study is the derivation of a (computationally) simple test deciding
in favour of $\mathcal{H}_{1}$ (resp. $\mathcal{H}_{0}$) when the
statistic $\Lambda(\hat{\bm{b}})$ is above (resp. below) the threshold
$\gamma$, and the quantizer design for each sensor (i.e. an optimized
$\tau_{k}$, $k\in\mathcal{K}$).

\section{Fusion Rules \label{sec:Rao Test}}

\subsection{Test derivation}

A common approach for composite hypothesis testing is given by the
GLRT \cite{Shoari2012}, whose expression is:
\begin{equation}
\Lambda_{\mathrm{G}}\triangleq2\,\ln[P(\hat{\bm{b}};\hat{\theta}_{1},\widehat{\bm{x}}_{T})\,/\,P(\hat{\bm{b}};\theta_{0})]\label{eq:GLR_general}
\end{equation}
where $P(\hat{\bm{b}};\theta,\bm{x}_{T})$ denotes the likelihood
as a function of $(\theta,\bm{x}_{T})$, whereas $\hat{\theta}_{1}$
and $\widehat{\bm{x}}_{T}$ are the \emph{Maximum Likelihood }(ML)\emph{
estimates} under $\mathcal{H}_{1}$ (i.e. $(\hat{\theta}_{1},\widehat{\bm{x}}_{T})\triangleq\arg\max_{(\theta,\bm{x}_{T})}P(\hat{\bm{b}};\theta,\bm{x}_{T})$).
It is clear from Eq. (\ref{eq:GLR_general}) that $\Lambda_{\mathrm{G}}$
requires the solution to an optimization problem. Unfortunately a
closed form for the pair $(\hat{\theta}_{1},\widehat{\bm{x}}_{T})$
is not available even for Gaussian noise. This increases the computational
complexity of its implementation, which typically involves a grid
approach on ($\theta,\bm{x}_{T}$), see e.g. \cite{Shoari2012}.

A different path for exploiting the two-sided nature of the problem
consists in adopting the rationale in \cite{Davies1987}. This allows
to extend score-based tests to the case of nuisance parameters present
solely under $\mathcal{H}_{1}$. Indeed, score-based tests require
the ML estimates of nuisances under $\mathcal{H}_{0}$ \cite{Kay1998},
which thus cannot be obtained, as they are not \emph{observable}.
The cornerstone of Davies' work is summarized as follows. If $\bm{x}_{T}$
were known in (\ref{eq:binary_test}), it would be easy to find a
simple test for a two-sided testing: indeed, in the latter case, the
Rao test seems a reasonable decision procedure \cite{Kay1998}. However,
since $\bm{x}_{T}$ is unknown in our setup, a \emph{family of statistics}
is instead obtained by varying $\bm{x}_{T}$. Thus, to overcome this
technical difficulty, Davies proposed the use of the \emph{maximum}
of the resulting family of the statistics, following a ``GLRT-like''
approach. In what follows, we will refer to the employed decision
test as \emph{Generalized Rao }(G-Rao)\emph{, }to underline the use
of Rao as the inner statistic employed in Davies approach, that is:
\begin{align}
\Lambda_{\mathrm{R}}\,\triangleq\, & \max_{\bm{x}_{T}}\left.\left(\frac{\partial\ln P(\hat{\bm{b}};\theta,\bm{x}_{T})}{\partial\theta}\right)^{2}\right|_{\theta=\theta_{0}}/\:\mathrm{I}(\theta_{0},\bm{x}_{T}),\label{eq:G-Rao_general}
\end{align}
where $\mathrm{I}(\theta,\bm{x}_{T})\triangleq\mathbb{E}\{\left(\frac{\partial\ln\left[P(\hat{\bm{b}};\theta,\bm{x}_{T})\right]}{\partial\theta}\right)^{2}\}$
is the \emph{Fisher Information }(FI) obtained assuming $\bm{x}_{T}$
is known, evaluated at $\theta_{0}$ in (\ref{eq:G-Rao_general}).
Our choice is motivated by reduced complexity of test implementation
(since $\hat{\theta}_{1}$ is not required, cf. Eq. (\ref{eq:G-Rao_general}),
and thus a grid implementation w.r.t. the sole $\bm{x}_{T}$ is required).

In order to obtain $\Lambda_{\mathrm{R}}$ explicitly, exploiting
the independence of sensors' measurements and reporting channels,
we expand $\ln\left[P(\hat{\bm{b}};\theta,\bm{x}_{T})\right]$ as:
\begin{gather}
\ln\left[P(\hat{\bm{b}};\theta,\bm{x}_{T})\right]=\sum_{k=1}^{K}\ln\left[P(\hat{b}_{k};\theta,\bm{x}_{T})\right]=\nonumber \\
\sum_{k=1}^{K}\{\hat{b}_{k}\,\ln\left[\alpha_{k}(\theta,\bm{x}_{T})\right]+(1-\hat{b}_{k})\,\ln\left[1-\alpha_{k}(\theta,\bm{x}_{T})\right]\}\label{eq:likelihood_error_prone}
\end{gather}
where $\alpha_{k}(\theta,\bm{x}_{T})\triangleq(1-P_{e,k})\beta_{k}(\theta,\bm{x}_{T})+P_{e,k}(1-\beta_{k}(\theta,\bm{x}_{T}))$
and $\beta_{k}(\theta,\bm{x}_{T})\triangleq F_{w_{k}}(\tau_{k}-\theta g(\bm{x}_{T},\bm{x}_{k}))$,
$F_{w_{k}}(\cdot)$ being the complementary cumulative distribution
function of $w_{k}$. On the other hand, the closed form of $\mathrm{I}(\theta,\bm{x}_{T})$
is \cite{Fang2013,Ciuonzo2013b}:
\begin{gather}
\mathrm{I}(\theta,\bm{x}_{T})=\sum_{k=1}^{K}\psi_{k}(\theta,\bm{x}_{T})\,g(\bm{x}_{T},\bm{x}_{k})^{2}\,,\label{eq:FI_error_prone}
\end{gather}
where
\begin{equation}
\psi_{k}(\theta,\bm{x}_{T})\triangleq\frac{(1-2\,P_{e,k})^{2}\:p_{w_{k}}^{2}(\tau_{k}-\theta g(\bm{x}_{T},\bm{x}_{k}))}{\alpha_{k}(\theta,\bm{x}_{T})\,\left(1-\alpha_{k}(\theta,\bm{x}_{T})\right)}\,.
\end{equation}
Plugging Eqs. (\ref{eq:likelihood_error_prone}-\ref{eq:FI_error_prone})
into (\ref{eq:G-Rao_general}), we obtain $\Lambda_{\mathrm{R}}$
explicitly as:
\begin{gather}
\Lambda_{\mathrm{R}}=\max_{\bm{x}_{T}}\frac{\left[\sum_{k=1}^{K}\nu_{k}(\hat{b}_{k})\,g(\bm{x}_{T},\bm{x}_{k})\right]^{2}}{\sum_{k=1}^{K}\psi_{k,0}\,g(\bm{x}_{T},\bm{x}_{k})^{2}},\label{eq: G-Rao explicit}
\end{gather}
where we have defined $\nu_{k}(\hat{b}_{k})\triangleq\frac{(1-2\,P_{e,k})\,p_{w_{k}}(\tau_{k})\,\left[\hat{b}_{k}-\alpha_{k,0}\right]}{\alpha_{k,0}(1-\alpha_{k,0})}$,
$\alpha_{k,0}\triangleq\alpha_{k}(\theta_{0},\bm{x}_{T})$ and $\psi_{k,0}\triangleq\psi_{k}(\theta_{0},\bm{x}_{T})$.
It is apparent that $\Lambda_{\mathrm{R}}$ (as well as $\Lambda_{\mathrm{G}}$)
is a function of $\tau_{k}$ (as $\widehat{\nu}_{k}(\hat{b}_{k})$
and $\psi_{k,0}$ \emph{both depend} on $\tau_{k}$), $k\in\mathcal{K}$,
(collected as $\bm{\tau}\triangleq\begin{bmatrix}\tau_{1} & \cdots & \tau_{K}\end{bmatrix}^{T}$)
which can be optimized to achieve improved performance. 

\subsection{Quantizer Design\label{subsec: Quantizer Design}}

It is worth noticing that (asymptotically-) optimal deterministic
quantizers cannot be obtained as in \cite{Fang2013,Ciuonzo2013b},
because no performance expressions are known in the literature for
tests based on the Davies approach \cite{Davies1987}. To this end,
we adopt a modified version of the rationale in \cite{Fang2013,Ciuonzo2013b}
and then we confirm its validity by simulations in Sec. \ref{sec: Simulation Results}.
Specifically, it is known that the (position $\bm{x}_{T}$) clairvoyant
Rao statistic $\bar{\Lambda}_{\mathrm{R}}$ (as well as the corresponding
clairvoyant GLR), is asymptotically (and assuming a weak signal\footnote{ That is $|\theta_{1}-\theta_{0}|=c/\sqrt{K}$ for some constant $c>0$
\cite{Kay1998}.}) distributed as \cite{Kay1998}
\begin{align}
\bar{\Lambda}_{\mathrm{R}}\overset{{\scriptstyle a}}{\sim} & \begin{cases}
\chi_{1}^{2} & \quad\mathrm{under}\quad\mathcal{H}_{0}\\
\chi_{1}^{'2}\left(\lambda_{Q}(\bm{x}_{T})\right) & \quad\mathrm{under}\quad\mathcal{H}_{1}
\end{cases},\label{eq:Asymptotic_performance}
\end{align}
where the non-centrality parameter $\lambda_{Q}(\bm{x}_{T})\triangleq(\theta_{1}-\theta_{0})^{2}\,\mathrm{I}(\theta_{0},\bm{x}_{T})$
(underlining dependence on $\bm{x}_{T}$) is given as:
\begin{align}
\lambda_{Q}(\bm{x}_{T}) & =\theta_{1}^{2}\sum_{k=1}^{K}\psi_{k,0}\,g(\bm{x}_{T},\bm{x}_{k})^{2}\,,\label{eq: non-centrality parameter}
\end{align}
with $\theta_{1}$ being the true value under $\mathcal{H}_{1}$.
Clearly the larger $\lambda_{Q}(\bm{x}_{T})$, the better the $\bm{x}_{T}-$clairvoyant
GLRT and Rao tests will perform when the target to be detected is
located at $\bm{x}_{T}$. Also, it is apparent that $\lambda_{Q}(\bm{x}_{T})$
is a function of $\tau_{k}$, $k\in\mathcal{K}$ (because of the $\psi_{k,0}$'s).
For this reason, with a slight abuse of notation we will use $\lambda_{Q}(\bm{x}_{T},\bm{\tau})$
and we choose the thresholds $\bm{\tau}$ to maximize $\lambda_{Q}(\bm{x}_{T},\bm{\tau})$,
that is $\bm{\tau}^{\star}\triangleq\arg\max_{\bm{\tau}}\;\lambda_{Q}(\bm{x}_{T},\bm{\tau})$.
In general, such optimization would lead to an optimized threshold
that will be dependent on $\bm{x}_{T}$ (and thus not practical).
However, for this specific problem the optimization can be decoupled
into the following set of $K$ independent threshold design problems,
which are \emph{independent of $\bm{x}_{T}$} (cf. Eq. (\ref{eq: non-centrality parameter})):
\begin{equation}
\arg\max_{\tau_{k}}\left\{ \psi_{k,0}(\tau_{k})=\frac{p_{w_{k}}^{2}(\tau_{k})}{\Delta_{k}+F_{w_{k}}(\tau_{k})\left[1-F_{w_{k}}(\tau_{k})\right]}\right\} \label{eq:objective_function_bsc}
\end{equation}
where $\Delta_{k}\triangleq[P_{e,k}\,(1-P_{e,k})]/(1-2P_{e,k})^{2}$.
It is known from the quantized estimation literature \cite{Papadopoulos2001,Rousseau2003}
that many unimodal and symmetric $p_{w_{k}}(\cdot)$'s with $\mathbb{E}\{w_{k}\}=0$
lead to $\tau_{k}^{\star}\triangleq\arg\max_{\tau_{k}}\psi_{k,0}(\tau_{k})=0$
(independent of $\Delta_{k}$); such examples are the Gaussian, Laplace,
Cauchy and the widely used generalized normal distribution (only in
the case $0\leq\epsilon\leq2$). Also, it has been shown in \cite{Ciuonzo2013b}
that $\tau_{k}=0$ is still a good (sub-optimal) choice even when
not corresponding to the optimizer for a specific noise pdf, especially
in the case of noisy ($P_{e,k}\neq0$) reporting channels. Therefore,
we employ $\tau_{k}=0$, $k\in\mathcal{K},$ in Eq.~(\ref{eq: G-Rao explicit}),
leading to the following further simplified expression for threshold-optimized
G-Rao test (denoted with $\Lambda_{\mathrm{R}}^{\star}$):
\begin{equation}
\Lambda_{\mathrm{R}}^{\star}\triangleq\max_{\bm{x}_{T}}\frac{4\left[\sum_{k=1}^{K}(1-2\,P_{e,k})\,p_{w_{k}}(0)\,g(\bm{x}_{T},\bm{x}_{k})\,(\hat{b}_{k}-\frac{1}{2})\right]^{2}}{\sum_{k=1}^{K}(1-2\,P_{e,k})^{2}\,p_{w_{k}}^{2}(0)\,g(\bm{x}_{T},\bm{x}_{k})^{2}}\label{eq:Rao_test_simplified_error_prone}
\end{equation}
which is considerably simpler than the GLRT, as it obviates solution
of a joint optimization problem w.r.t. $(\bm{x}_{T},\theta)$ (which
depends on $p_{w_{k}}(\cdot)$). Furthermore, the corresponding optimized
non-centrality parameter, denoted with $\lambda_{Q}^{\star}(\bm{x}_{T})$,
is given by:
\begin{equation}
\lambda_{Q}^{\star}(\bm{x}_{T})\triangleq4\theta_{1}^{2}\sum_{k=1}^{K}\left[(1-2\,P_{e,k})^{2}\,p_{w_{k}}^{2}(0)\,g(\bm{x}_{T},\bm{x}_{k})^{2}\right].\label{eq:optimized-noncentralitypar-BSC-1}
\end{equation}

\subsection{Computational Complexity}

As detailed in \cite{Niu2006b,Shoari2012,Ciuonzo2016}, the GLRT is
usually implemented by means of a grid discretization. More specifically,
assuming that $\bm{x}_{T}$ and $\theta$ belong to limited sets $\mathrm{S}_{\bm{x}_{T}}\subset\mathbb{R}^{d}$
and $S_{\mathrm{\theta}}\subset\mathbb{R}$, respectively, the search
space ($\bm{x}_{T},\theta)$ required for (\ref{eq:GLR_general})
is then discretized into: ($a$) $N_{\bm{x}_{T}}$ position bins in
$\mathbb{R}^{d}$, each one associated to a center bin position, say
$\bm{x}_{T}[i]$, $i\in\{1,\ldots N_{\bm{x}_{T}}\}$; ($b$) $N_{\theta}$
amplitude bins in $\mathbb{R}$, each one to associated to a center
bin amplitude, say $\theta[j]$, $j\in\{1,\ldots N_{\theta}\}$. Similarly,
the G-Rao statistic is implemented by discretizing the sole search
space of $\bm{x}_{T}$, leading to:
\begin{gather}
\Lambda_{\mathrm{R}}\approx\max_{i=1,\ldots N_{\bm{x}_{T}}}\frac{\left[\sum_{k=1}^{K}\nu_{k}(\hat{b}_{k})\,g(\bm{x}_{T}[i],\bm{x}_{k})\right]^{2}}{\sum_{k=1}^{K}\psi_{k,0}\,g(\bm{x}_{T}[i],\bm{x}_{k})^{2}}\,.\label{eq: G-Rao discretization}
\end{gather}
Thus, its complexity is $\mathcal{O}\left(K\,N_{\bm{x}_{T}}\right)$,
thus providing a \emph{significant complexity reduction} w.r.t. the
GLR, as reported in Tab.~\ref{tab: Detectors comparison}.
\begin{table}
\begin{centering}
\medskip{}
\par\end{centering}
\centering{}\caption{Complexity comparison of decision statistics.\label{tab: Detectors comparison}}
\begin{tabular}{c|c}
\hline 
\noalign{\vskip0.1cm}
\multicolumn{1}{|c|}{\textbf{Fusion Rule}} & \multicolumn{1}{c|}{\textbf{Computational Complexity}}\tabularnewline[0.1cm]
\hline 
\hline 
\noalign{\vskip0.1cm}
GLR & $\mathcal{O}\left(K\,N_{\bm{x}_{T}}\,N_{\theta}\right)$ (Grid search)\tabularnewline[0.1cm]
\hline 
\noalign{\vskip0.1cm}
G-Rao &  $\mathcal{O}\left(K\,N_{\bm{x}_{T}}\right)$ (Grid search)\tabularnewline[0.1cm]
\hline 
\end{tabular}
\end{table}
 
\begin{figure}[t]
\centering{}\includegraphics[width=1\columnwidth]{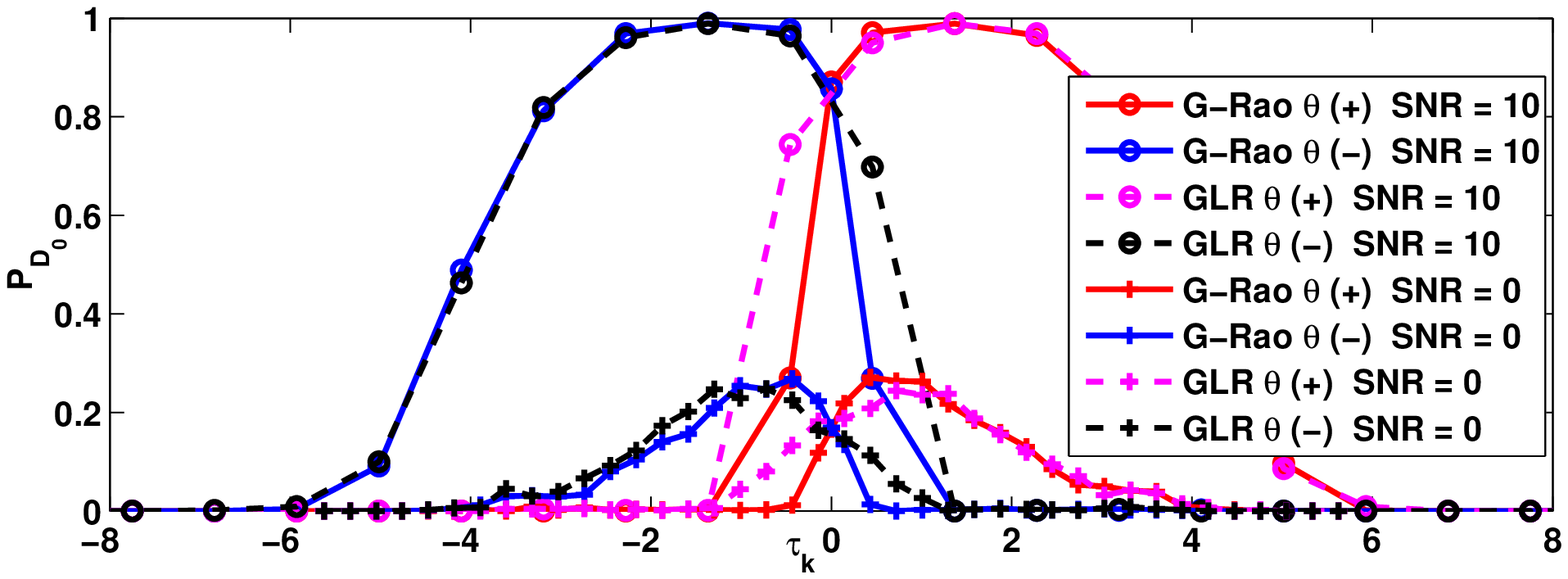}\caption{$P_{D_{0}}$ vs $\tau_{k}=\tau$, $P_{F_{0}}=0.01$; WSN with $K=49$
sensors, $P_{e,k}=0$, $\mathrm{SNR}\in\{0,10\}$ (amplitude signal
with positive/negative polarity).\label{fig: Pdo vs tau}}
\end{figure}
\begin{figure}[t]
\centering{}\includegraphics[width=1\columnwidth]{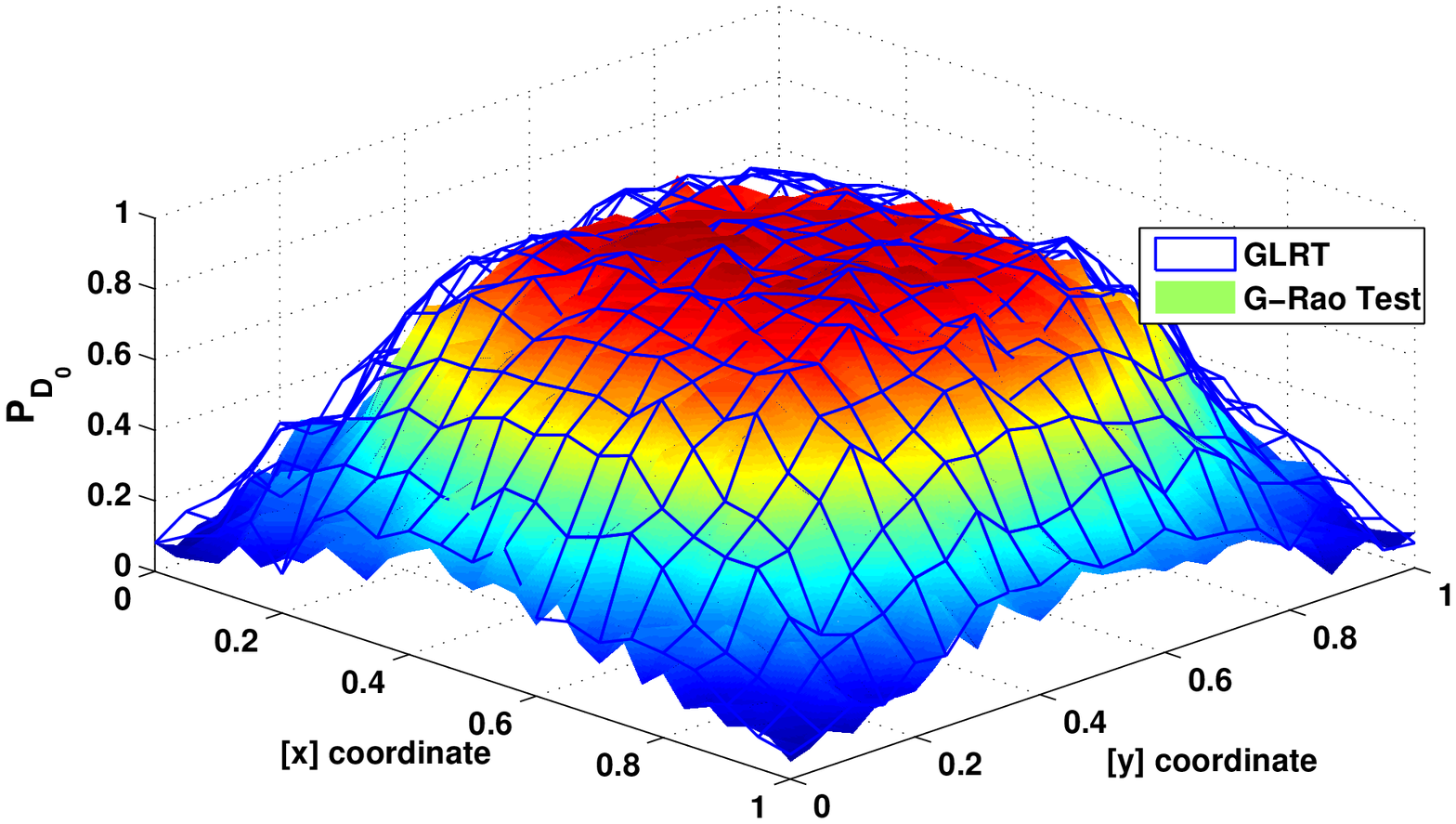}\caption{$P_{D_{0}}$ vs $\bm{x}_{T}$, $P_{F_{0}}=0.01$; WSN with $K=49$
sensors, $\tau_{k}=0$, $P_{e,k}=0$, $\mathrm{SNR=5\,\mathrm{dB}}$.\label{fig: Pdo vs xt (tau =00003D 0) Pe =00003D 0}}
\end{figure}
\begin{figure}[t]
\centering{}\includegraphics[width=1\columnwidth]{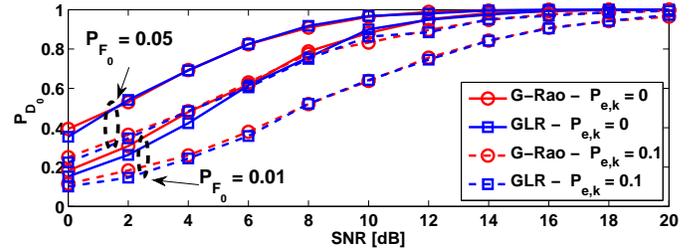}\caption{$P_{D_{0}}$ vs. $\mathrm{SNR}$ ($\mathrm{dB}$), $P_{F_{0}}\in\{0.05,0.01\}$;
WSN with $K=49$ sensors, $\tau_{k}=0$, $P_{e,k}=P_{e}\in\{0,0.1\}$.\label{fig: Pdo vs SNR (GLR vs GRao)}}
\end{figure}

\section{Gaussian noise Analysis\label{sec: Simulation Results}}

In this section we compare G-Rao and GLR tests, by evaluating their
performance in terms of system false alarm and detection probabilities,
defined as $P_{F_{0}}\triangleq\Pr\{\Lambda>\gamma|\mathcal{H}_{0}\}$
and $P_{D_{0}}\triangleq\Pr\{\Lambda>\gamma|\mathcal{H}_{1}\}$, respectively,
where $\Lambda$ is the statistic employed at the FC. Additionally,
we will validate the zero-threshold choice obtained in Sec. \ref{subsec: Quantizer Design}. 

To this end, we consider a 2-D scenario ($\bm{x}_{T}\in\mathbb{R}^{2}$)
where a WSN composed of $K=49$ sensors is employed to detect the
presence of a target within the (square) region $\mathcal{A}\triangleq[0,1]^{2}$,
being the surveillance area. For simplicity the sensors are arranged
according to a regular square grid covering $\mathcal{A}$. With reference
to the sensing model\footnote{To complement our analysis, this letter provides corresponding results
for Laplace noise in next section.}, we assume $w_{k}\sim\mathcal{N}(0,\sigma_{w}^{2})$, $k\in\mathcal{K}$
(also w.l.o.g. we set $\sigma_{w}^{2}=1$). Also, the AAF chosen is
$g(\bm{x}_{T},\bm{x}_{k})\triangleq1\,/\,\sqrt{1+\left(\left\Vert \bm{x}_{T}-\bm{x}_{k}\right\Vert /\,\eta\right)^{\alpha}}$
(i.e. a power-law), where we have set $\eta=0.2$ (viz. approximate
target extent) and $\alpha=4$ (viz. decay exponent). Finally, we
define the target Signal-To-Noise Ratio (SNR) as $\mathrm{SNR}\triangleq10\,\log_{10}(\theta^{2}/\sigma_{w}^{2})$.
Initially, we assume ideal BSCs, i.e. $P_{e,k}=0$, $k\in\mathcal{K}$.

As explained before, $\Lambda_{\mathrm{G}}$ and $\Lambda_{\mathrm{R}}$
are implemented by means of grids for $\theta$ and $\bm{x}_{T}$.
Specifically, the search space of the target signal $\theta$ is assumed
to be $S_{\theta}\triangleq\left[-\bar{\theta},\bar{\theta}\right]$,
where $\bar{\theta}>0$ is such that the $\mathrm{SNR}=20\,\mathrm{dB}$.
The grid points are then chosen as $\begin{bmatrix}-\bm{g}_{\theta}^{T} & 0 & \bm{g}_{\theta}^{T}\end{bmatrix}^{T},$
where $\bm{g}_{\theta}$ collects target strengths corresponding to
the SNR dB values $-10:1:20$ (thus $N_{\theta}=63$). Differently,
the search space of the target position $\bm{x}_{T}$ is (naturally)
assumed to coincide with the surveillance area, i.e. $\mathrm{S}_{\bm{x}_{T}}=\mathcal{A}$.
The 2-D grid points are then obtained by regularly sampling $\mathcal{A}$
with $N_{\bm{x}_{T}}=N_{c}^{2}$ points, where $N_{c}=100$.

First, in Fig. \ref{fig: Pdo vs tau} we show $P_{D_{0}}$ (under
$P_{F_{0}}=0.01$) versus a common threshold choice for all the sensors
$\tau_{k}=\tau$, $k\in\mathcal{K}$, for a target whose location
is randomly drawn according to a uniform distribution within $\mathcal{A}$.
It is apparent that in the low-SNR limit $\tau=0$ represents a nearly-optimal
solution, since the optimal value of $\tau$ found numerically depends
on the polarity of $\theta$, which is unknown. This both applies
to GLR and G-Rao as well. Secondly, in Fig. \ref{fig: Pdo vs xt (tau =00003D 0) Pe =00003D 0},
we report $P_{D_{0}}$ (under $P_{F_{0}}=0.01$) versus target location
$\bm{x}_{T}$ (for $\mathrm{SNR}=5\,\mathrm{dB}$), in order to obtain
a clear comparison of detection performance over the entire surveillance
area $\mathcal{A}$. It is apparent that the G-Rao test presents only
marginal loss over the GLRT. Additionally the $P_{D_{0}}(\bm{x}_{T})$
profile is qualitatively similar for both rules, and underlines lower
detection performance at the boundaries of the surveillance area.
This can be attributed to regular displacement of the WSN within $\mathcal{A}$.
Finally, in Fig. \ref{fig: Pdo vs SNR (GLR vs GRao)} we compare the
$P_{D_{0}}$ (for $P_{F_{0}}\in\{0.05,0.01\}$) of considered rules
(for a target with randomly drawn position within $\mathcal{A}$)
versus $\mathrm{SNR}$ ($\mathrm{dB}$), in order to obtain a comparison
of detection sensitivity versus the signal strength. It is apparent
that both rules perform very similarly over the whole SNR range, as
well as for a different quality of the reporting channel ($P_{e,k}=P_{e}\in\{0,0.1\}$).

\section{Laplace noise Analysis \label{sec: Laplace results}}

In this section the focus will be on $w_{k}$s modelled as \emph{Laplace
noise}. Similarly, we will validate the zero-threshold choice proposed
in the paper also for this case. With reference to the sensing model,
we assume $w_{k}\sim\mathcal{L}(0,\beta_{k})$, $k\in\mathcal{K}$
(here $\mathcal{L}(\mu,\beta)$ is used to denote a Laplace pdf with
mean $\mu$ and scale parameter $\beta$). Also for simplicity, we
assume that each $\beta_{k}$ is chosen such that $\mathrm{\mathbb{E}}\{w_{k}^{2}\}=1$.
Furthermore, we define the target Signal-To-Noise Ratio (SNR) as $\mathrm{SNR}\triangleq10\,\log_{10}(\theta^{2}/\mathrm{\mathbb{E}}\{w_{k}^{2}\})$.
Initially, we assume ideal BSCs, i.e. $P_{e,k}=0$, $k\in\mathcal{K}$.
Finally, we remark that we use the same grid implementation of GLRT
and G-Rao test employed in the previous section for Gaussian noise.

\begin{figure}
\centering{}\includegraphics[width=1\columnwidth]{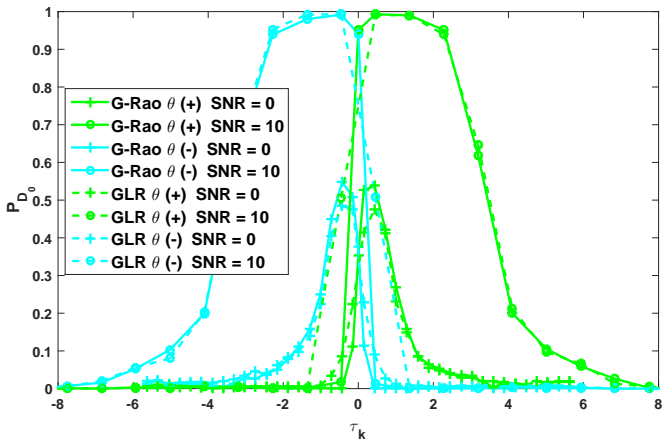}\caption{$P_{D_{0}}$ vs $\tau_{k}=\tau$, $P_{F_{0}}=0.01$; WSN with $K=49$
sensors, $P_{e,k}=0$, $\mathrm{SNR}\in\{0,10\}$ (amplitude signal
with positive/negative polarity).\label{fig: Pdo vs tau-Laplace}}
\end{figure}

First, in Fig. \ref{fig: Pdo vs tau-Laplace} we show $P_{D_{0}}$
(under $P_{F_{0}}=0.01$) versus a common threshold choice for all
the sensors $\tau_{k}=\tau$, $k\in\mathcal{K}$, for a target whose
location is randomly drawn according to a uniform distribution within
$\mathcal{A}$. It is apparent that in the low-SNR limit $\tau=0$
represents a nearly-optimal solution, since the optimal value of $\tau$
found numerically depends on the polarity of $\theta$, which is unknown
(this both applies to GLR and G-Rao as well). Similar results have
been observed also in the case of Gaussian noise in the paper itself.

\begin{figure}
\centering{}\includegraphics[width=1\columnwidth]{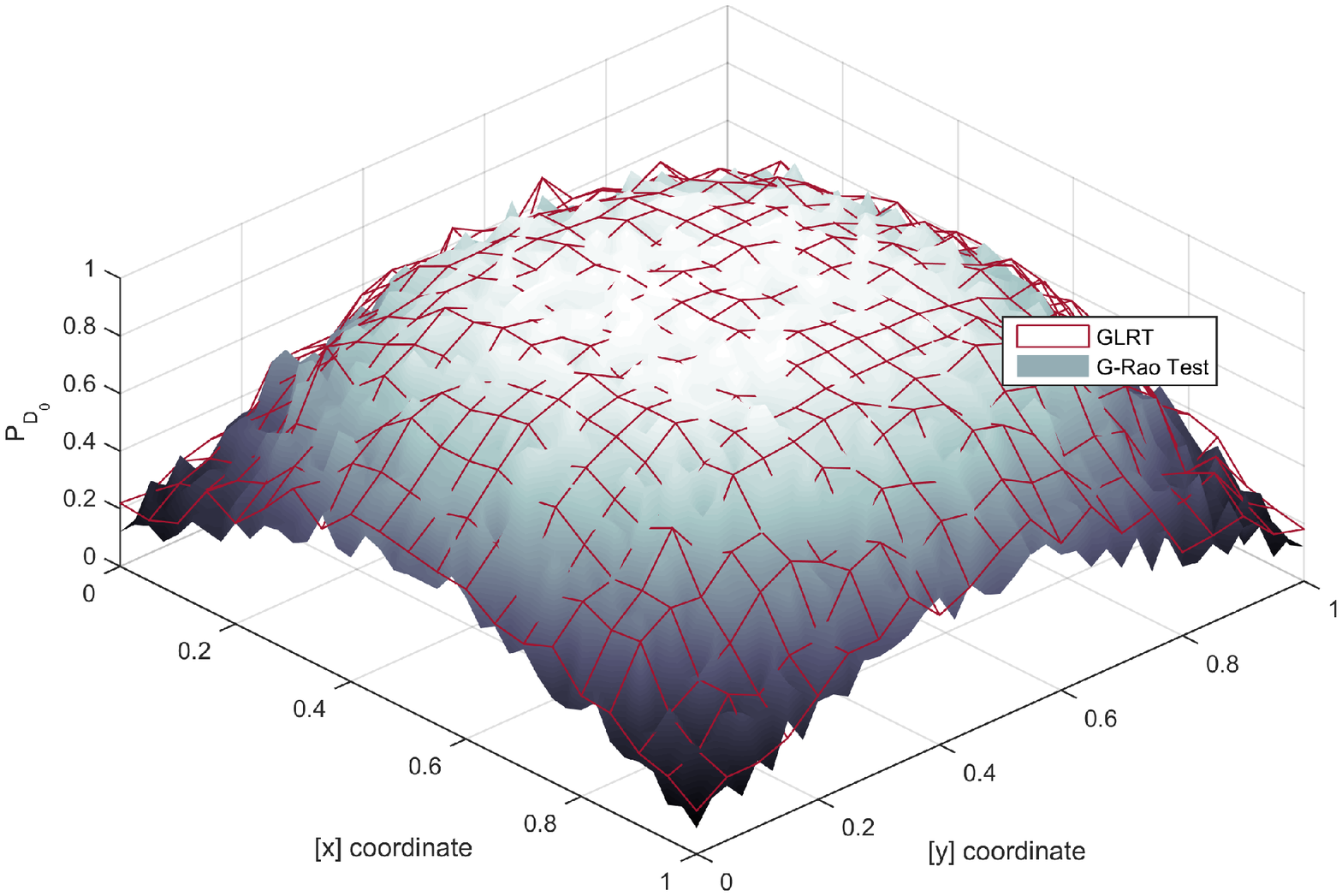}\caption{$P_{D_{0}}$ vs $\bm{x}_{T}$, $P_{F_{0}}=0.01$; WSN with $K=49$
sensors, $\tau_{k}=0$, $P_{e,k}=0$, $\mathrm{SNR=5\,\mathrm{dB}}$.\label{fig: Pdo vs xt (tau =00003D 0) Pe =00003D 0 Laplace}}
\end{figure}

Secondly, in Fig. \ref{fig: Pdo vs xt (tau =00003D 0) Pe =00003D 0 Laplace},
we report $P_{D_{0}}$ (under $P_{F_{0}}=0.01$) versus target location
$\bm{x}_{T}$ (for $\mathrm{SNR}=5\,\mathrm{dB}$), in order to obtain
a clear comparison of detection performance over the entire surveillance
area $\mathcal{A}$. It is apparent that the G-Rao test presents only
marginal loss over the GLRT. By looking at the similar qualitative
behaviour between Laplace and Gaussian noise (reported in the paper),
we conclude that such trend is quite general for unimodal zero-mean
noise pdfs.

\begin{figure}
\centering{}\includegraphics[width=0.9\columnwidth]{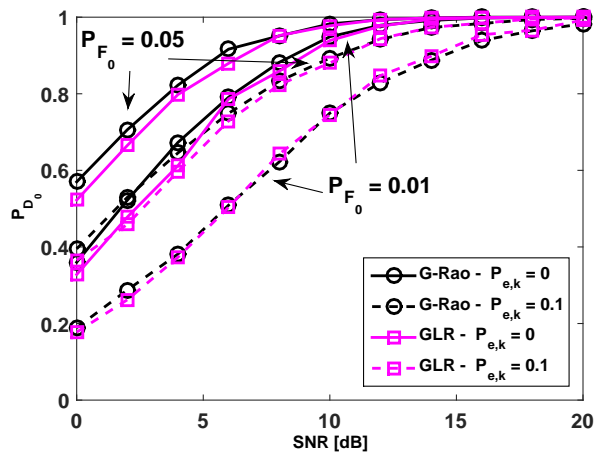}\caption{$P_{D_{0}}$ vs. $\mathrm{SNR}$ ($\mathrm{dB}$), $P_{F_{0}}\in\{0.05,0.01\}$;
WSN with $K=49$ sensors, $\tau_{k}=0$, $P_{e,k}=P_{e}\in\{0,0.1\}$.\label{fig: Pdo vs SNR (GLR vs GRao) Laplace}}
\end{figure}

Finally, in Fig. \ref{fig: Pdo vs SNR (GLR vs GRao) Laplace} we compare
the $P_{D_{0}}$ (for $P_{F_{0}}\in\{0.05,0.01\}$) of considered
rules (for a target with randomly drawn position within $\mathcal{A}$)
versus $\mathrm{SNR}$ ($\mathrm{dB}$), in order to obtain a comparison
of detection sensitivity versus the signal strength. It is apparent
that both rules perform very similarly over the whole SNR range, as
well as for a different quality of the reporting channel ($P_{e,k}=P_{e}\in\{0,0.1\}$),
with G-Rao slightly outperforming the GLRT at low SNR.

\section{Conclusions\label{sec: Conclusions}}

We developed a generalized version of the Rao test (G-Rao, based on
\cite{Davies1987}) for decentralized detection of a non-cooperative
target emitting an unknown deterministic signal ($\theta$) at unknown
location ($\bm{x}_{T}$), as an attractive (low-complexity) alternative
to GLRT (the latter requiring a grid search on the whole space $(\theta,\bm{x}_{T})$)
for a general model with quantized measurements, zero-mean, unimodal
and symmetric noise (pdf), non-ideal and non-identical BSCs. Since
$\bm{x}_{T}$ is a nuisance parameter present only under $\mathcal{H}_{1}$
(i.e. when $\theta\neq0$), the G-Rao statistic arises from maximization
(w.r.t. $\bm{x}_{T}$) of a family of Rao decision statistics, obtained
by assuming $\bm{x}_{T}$ known. We also developed a reasonable criterion
for optimized sensor thresholds: the zero choice was shown to be appealing
for many pdfs of interest. This result was exploited to optimize the
performance of G-Rao and GLR tests. Also, it was shown through simulations
that the G-Rao test, achieves practically the same performance as
the GLRT in the cases considered.

\bibliographystyle{IEEEtran}
\bibliography{IEEEabrv,sensor_networks}

\end{document}